# A Survey on Rural Internet Connectivity in India

Shruthi K.A., G.V. Ihita, *Student Member, IEEE*, and Sachin Chaudhari, *Senior Member, IEEE*

*Abstract*—Rural connectivity is widely research topic for several years. In India, around 70% of the population have poor or no connectivity to access digital services. Different solutions are being tested and trialled around the world, especially in India. They key driving factor for reducing digital divide is exploring different solutions both technologically and economically to lower the cost for the network deployments and improving service adoption rate. In this survey, we aim to study the rural connectivity use-cases, state of art projects and initiatives, challenges, and technologies to improve digital connectivity in rural parts of India. The strengths and weakness of different technologies which are being tested for rural connectivity is analyzed. We also explore the rural use-case of 6G communication system which would be suitable for rural Indian scenario.

## I. INTRODUCTION

Although India has already made the giant leap in the digital revolution, there is still a long way to connect the unconnected. As per a report by the Telecom Regulatory Authority of India (TRAI), there is a 93% broadband penetration in urban India versus a mere 29.3% in rural India [1], [2]. Exacerbated by COVID-19, this digital divide is worsening the existing societal and economic inequalities in the country. Further, there are concerns about widening the divide due to adopting new digital developments such as digital wallets, IoT, and AI.

There is a need to empower the rural community members through reliant connectivity and provide them with the relevant skill sets to enjoy the benefits which are currently spread unevenly across urban and rural India [3]. Leveraged and scaled, the power of information exchange can provide various opportunities such as smart agriculture, education, telemedicine services, ease of transfer of payments through government schemes, e-commerce, awareness on urban developments, and remote job opportunities [4].

The major challenges that exist when increasing internet penetration in India are unreliable electricity, affordability of the connectivity and user equipment, negligible industry incentives for low-cost ICT, topographical challenges, cost of infrastructure, backhaul issues, and the performance trade-offs [5], [6]. Further, the lack of digital awareness in using technology and the cultural reservations as well as language causing the digital divide a huge challenge in India [7]. The key solution lies in formulating strategies and initiatives which require inputs from operators, equipment manufacturers, academia, government, and civil bodies to understand the requirements of the rural communities to create sustainable solutions for them to leverage for development [8].

There is extensive research done in solving the digital divide problem around the world for the Indian scenario [3]–[13]. In [3], the authors state the need for innovations in the business models for rural areas along with the innovations in the technology. In [4], the TRAI and the researchers in [5], [6] highlights the different challenges, uses-cases, different solutions and need for innovation in business models for using 5G technology in the Indian setting. In [7], the authors presents the techno-economic feasibility of rural solutions in India. In [8], the Department of Telecommunication (DoT), India presents research on 5G technology and its uses for the Indian market with slight focus on rural areas connectivity. A study on role of telecommunication in managing pandemic such as COVID-19 has highlighted the importance of digital connectivity especially in rural areas [9]. The authors in [10] explore low cost solutions using TVWS in providing rural connectivity in India. 5G encompass combination of various technologies such as 5G, network slicing, TVWS, and WiFi integration, which are being tested to solve rural connectivity challenge [11]. The need for innovative business models for rural scenarios with MNO and business ecosystem are discussed in [12] whereas the benefits of using 5G NHN in the Indian scenario is explored in [13] which shows that it is cost efficient for rural setting. The need for supportive policies for policies for encouraging NHN [14] and 5G deployments in rural areas with special focus on India scenario is discussed in [15].

Our research is different from the existing research presented in [3]–[13]. The authors in [3], [12] present a generic study on rural digital connectivity and possible solutions but not specific to India. In [6] the study gives an glimpse of the rural connectivity solutions in India only using fiber backhaul, TVWS and WiFi. In [7], [9], [11], [13] are use of specific solutions rather than an overview of all the possible solutions of rural digital connectivity. In [4], [5], [8], [14], [15] include different possibilities and policies that could solve the issues and the roles of stakeholders but not an extensive study on suitability to the address the challenges and solutions for the Indian scenario.

According to researchers in [16], [17], lack of awareness of appropriate solutions for digital connectivity in India is a major roadblock that should be addressed. Also, the research performed by [6] shows that around 70% of surveyed institutions didn't know the existence of BharatNet. The key motivation for this survey is to overcome the lack of awareness among the people regarding the topics such as initiatives accessible

Shruthi K A, is with the StrathSDR Lab, University of Strathclyde, Glasgow, G1 1XQ Scotland, UK (e-mail: k.a.shruthi@strath.ac.uk)

G.V. Ihita and Sachin Chaudhari are with the Signal Processing and Communication Research Centre (SPCRC), International Institute of Information Technology, Hyderabad, 500032, India (e-mail: ihita.g@iiit.ac.in, sachin.chaudhari@iiit.ac.in).

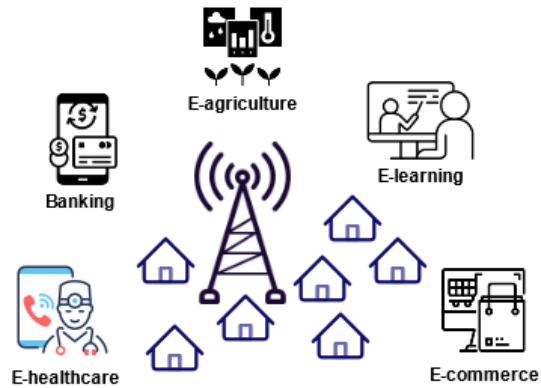

Fig. 1. Rural connectivity use cases

for rural digital connectivity, network planning based on the possible use-cases in rural areas rather than a generic urban style network, the key challenges that the deployment needs to address in the rural areas, possible solutions, and its potential. In this research, a survey on rural internet connectivity for an Indian scenario is presented. In this paper, we look at the following and try to answer the questions about:

- What are the key use-cases of rural digital connectivity?
- What are the current state-of-the-art initiatives and project undertaken to improve rural connectivity in India?
- What are the major challenges that needs to be addressed to improve rural telecommunication?
- What are the tests and the trials around the world for improving rural connectivity? Have these solutions been tested in India?
- How are 6G solutions going to address rural connectivity needs?

This paper is organised as follows: Section II presents the rural connectivity use-cases and Section III the current state-of-the-art initiatives and projects to improve rural connectivity in India. Section IV presents the main challenges which has to be solved for reducing digital divide. Section V presents a survey on different technologies such as 5G, TVWS, and satellite internet, their role in solving the issues and the trials undertaken around the world. Furthermore, there is brief discussion on 6G solutions which would be focusing on rural connectivity. Finally, conclusions are drawn in Section VI.

## II. RURAL CONNECTIVITY USE-CASES

The internet provides immense potential for the development of rural communities. The prominent and emerging use-cases for digital connectivity in rural India, as shown in Figure 1, are explained below:

1) **Agriculture**: The main occupation of rural communities in India is agriculture. Over the years, a declining employment size of the agricultural sector has been observed. The internet can improve agricultural productivity such as for sharing essential information on sowing, crop protection, improving soil fertility and weather. Further, Early-Warning-Systems (EWS) provide disaster management support helping farmers prepare for sporadic events such as floods, drought, or even pest and disease outbreaks, thus preventing significant crop loss. It allows for a structured distribution of subsidies such as fertilizers. With access to internet connectivity farmers can get access to trusted channels where they can receive best market price of their commodity and other daily updates for their produce to ensuring they receive fair returns. The National Agriculture Market or e-Nam, an electronic trading portal, [18] brings regularisation on how crops and production are distributed across the country along with quality testing mechanism with govt standard and farmer access to a bigger market place. Internet helps farmers manage livestock. It enables the improvement of various aspects of livestock production and education.

2) **Healthcare**: Monitoring health and remotely accessing healthcare services by rural or isolated communities can be achieved through reliant connectivity. Telehealth includes long-distance clinical health care, patient and professional health-related education, public health, and health administration. It has been used by emergency medical personnel and for consultation during natural disasters and in military battle situations. The National Health Portal [19], initiative of the Ministry of Health and Family Welfare, Government of India aims to cover services such as online medical consultation, online medical records, online medicine supply management and pan-India exchange for patient information. Mobile applications such as Pradhan Mantri Surakshit Matritva Abhiyan [20] for pregnancy care, NHP Directory Services [21] for information on hospitals across India and Indradhanush Immunization [22] to track immunization status of children are helping with ease of access of healthcare services.

3) **Education**: Connecting rural communities help enable distance education in remote parts of the country, including vocation training classes to build skill-sets such as tailoring and weaving. Digitizing and recording teaching material for students with visual impairments is making the education sector even more inclusive. There are various initiatives under the programme, National Mission on Education through Information and Communication Technology (NMEICT) [23] such as National Digital Library (NDL) [24] which provides a virtual learning resource repository and SWAYAM Prabha [25] initiative providing 32 High Quality Educational Channels through DTH (Direct to Home). Further, the virtual labs [26] initiative of NMEICT provides remote access to Labs in various disciplines of science and engineering to support learning basic and advanced concepts through remote experimentation. Professor Raj Reddy center on Technology in Service of Society [27]at IIIT Hyderabad has multiple initiatives targeting rural youth education and teacher training. The centre is supporting various initiatives including the establishment of a learning resource/platform for underprivileged students with access to a non-smart mobile phone(feature phones) and voice network. They are also establishing virtual labs to provide remote access to physical labs and research centres.

4) **E-governance**: Various government services related to jobs, education, wellness, pension, justices are available on the National Government Services Portal [28]. Further other pandemic based essential services such as information on the vaccination centre are available on this platform. Connectivity is critical to rural sector to easily avail these e-services. [29], [30].

5) **Banking**: Reliant and secure digital connectivity can strengthen rural access to financial services enabling them to find appropriate insurance, support to manage and mitigate risk. It will assist with better farmer profiling systems for agricultural credit decisions and protection for farmers in times of bad weather or disaster. Governments also use the internet to increase their outreach, improve understanding of policies, and expand credit penetration. Under the jurisdiction of the Ministry of Finance, the National Bank for Agriculture and Rural Development(NABARD) [31] provides overall regulation and licensing of regional rural banks. The Pradhan Mantri Jan Dhan Yojana [32] aims to expand affordable access to financial services such as bank accounts, remittances, credit, insurance, and pensions. The Government of India scheme, Kisan Credit Card(KCC), also aims to provide credit in activities like the cultivation of crops, post-harvesting, expenditure, maintenance cost for assets used in agriculture, and investment requirements in agriculture and related activities. Further, the Indian government is promoting e-wallets and has launched many UPI (united payment interface) solutions and BHIM app for a smooth transition to digital payments.

6) **E-commerce**: E-commerce opportunities have been crucial in purchasing and selling agro-based products and handicrafts. With connectivity, comes access to regional and global markets. Rural communities have a platform to showcase diverse cultures and crafts promoting tourism. Indian government initiative called the GeM portal is an e-marketplace from where everyday common consumer goods can be procured. This platform allows on-board products of self-help groups, tribal communities, artisans, weavers etc, helping them flourish their businesses and maintain market transparency. Handicrafts and paintings from India, such as Kalamkari paintings, Lepakshi handicrafts, and Kondapalli toys, can be displayed and sold in international markets via e-commerce platforms.

III. STATE-OF-THE-ART INITIATIVES FOR IMPROVING RURAL CONNECTIVITY

The government of India is trying to address the issues for rural internet penetration using different projects and initiatives such as BharatNet, PMWani, Digital India, and an end-to-end 5G test bed. [6], [8].

1) **Digital India Initiatives**: Launched by the Government of India, the Digital India campaign aims to provide services of the Government available digitally to all citizens [4], [8] . They aim to connect rural areas with high-speed internet networks and improve digital literacy via stable and secure digital infrastructure. Various programs are catering to the infrastructure and service needs. Digital infrastructure contributors are:
   - BBNL: Optical fibre based broadband connectivity to Gram Panchayats.
   - CERT-IN: Requirement of contribution to create a secure cyberspace
   - Common Services Centres (CSCS): CSCS are access points for delivery of various electronic services to villages in India

   Connectivity infrastructure [33] connecting urban and rural teachers/students using low cost-affordable access cum computing devices provide on-line testing and certification.

2) **BharatNet**: Earlier called the National Optic Fiber Network (NOFN), Bharat Net mission implemented by Bharat Broadband Network Limited aims to create a robust middle-mile infrastructure using optical fiber for reaching broadband connectivity to all Gram Panchayats in India [34]. It also aims at connecting all of India's households, specifically rural families, through affordable high-speed internet connectivity. As per the project, there are 1,59,871 Gram Panchayats to which optical fiber cable has been connected and necessary equipment installed.

3) **AirJaldi**: AirJaldi is a social enterprise based in India aiming to provide wireless broadband connectivity to remote rural areas at reasonable rates. This includes,
   - Internet access in tough terrains of Northern India
   - Build and own infrastructure for internet distribution
   - Implement a business model where risks and returns of running rural networks are shared with local entrepreneurs.

4) **PM Wani (Wi-Fi Access Network Interface) scheme**: As per the scheme, a decentralized system of public access points will be implemented to increase internet connectivity across the country to address last-mile connectivity at cheap rates. There is no license, no registration, and no fee required for the PDOs (Public Data Offices), which could be small shops or even Common Service Centres as access points. [35].
5) **OceanNet** : The low cost internet connectivity solution can extend internet signals to fishermen up to 60 km at sea and 45 km beyond the range of cell phone towers [16], [17]. It is particularly essential for real time alert communications and disaster warnings.

The critical challenge of reliable power and electricity needs to be addressed. The government of India launched the Deendayal Upadhyaya Gram Jyoti Yojana for village electrification and providing electricity distribution infrastructure in the rural areas.

## IV. CHALLENGES SPECIFIC TO INDIAN SETTING

In this section, the different challenges in providing rural internet connectivity is studied. Though the government is encouraging initiatives and projects to improve rural connectivity, there is to understand the depth of challenges in providing last-mile connectivity and scale the network. In [4], [6], [7], [17], the authors describe different challenges related to rural internet connectivity in India. Furthermore, the key challenges as described below with respect to costs, ARPU, digital awareness, power, spectrum bands, and terrain, needs to be addressed to tackle this issue.

1) **Topography and population distribution**: India has diverse topography ranging plane, plateau to mountain and desserts and extreme climatic conditions. The topography of the village and its population distribution plays a crucial role in the internet connectivity. A sparsely populated village is given lowest priority while deploying networks. An uneven terrain poses a challenge in providing digital connectivity [6], [36]. A smooth terrain with fewer obstructions are favourable for providing rural connectivity. Therefore, a unique solution is not possible for rural connectivity. The population of villages in rural parts of India can vary from as low as 100 people per village to more than 10,000 people per village [1]. Hence, population distribution of the village needs to be one of the key factor in demands estimations and network planning.
2) **Cost**: The main factor which makes rural connectivity difficult is cost of providing internet solutions in rural areas. As stated by authors in [4], [13], [37], the high cost of network is due to various factors such as non-existence of either backhaul or close by point of presence (PoP), cost of equipment, repair and maintenance, and spectrum licensing in hard to reach locations. By lower the cost of rural deployments, the scalability of the networks increase. The techno-economic study of different technologies highlight the need for reduced deployment and running costs in rural areas for improving rural connectivity [7], [13], [38]–[40]. Furthermore, [4], [6], [41], [42] shows that provisioning 2G, 3G, 4G, or 5G in rural regions will be difficult unless the network's cost is reduced. The economic cost modelling used in urban and industrial sectors is not appropriate for rural locations.
3) **Per-capita income**: Generally, the people living in rural areas have lower per-capita income compared to urban areas. The pricing should be attractive for people living in rural areas to spend on digital applications [1], [2], [4], [6], [13]. There is a need for innovation in lower costs and new revenue models to attract rural areas customers to enrol for internet services. The pricing of services could be different in rural areas to attract in rural subscribers.
4) **Business models**: The lack of innovative business models is the major challenge for rural connectivity [2], [6], [7], [43]. The traditional business models are not suitable for rural scenarios as these focuses on high ARPU, high customer base, and high investment duration [3]. In rural scenarios, the traffic generated is low and it is a loss making business. This fuels the need to find innovation in business models to convert this business to profit making. The business model needs to focus on factors such as low ARPU, low to moderate customer base, and moderate investment duration [44].
5) **Funding and investment**: The MNOs and ISPs build network with future profits in mind. They obtain funding and investment from big banks and government agencies [4]. As the rural telecommunication business is non profitable for a new entrant or InP, the lack of funding allocated for rural last-mile connectivity and its development is another major roadblock [6]. There is a need for innovations in the possible funding model for local operators in rural areas. The investment requires for setting a telecommunication service is significant and to make it sustainable is key for scalability of any rural connectivity solutions.
6) **Technology**: The mobile communication technology needs to be specifically modified for meeting the rural communities' requirements. The key focus of rural networks are on factors such as high coverage (up to 10 km radius), high data rates (100 Mbps or more), high energy efficiency, high spectral efficiency, support all use-cases using single network and smooth roaming [2], [4], [6], [13], [29], [45]. There is also a need for local spectrum licensing to encourage micro-operators or infrastructure providers (InP) [12], [46]. Open network technologies such as OpenRAN, 5G open source base stations and other open source technologies are widely researched around the world which is expected to significantly reduce the cost of deployments [47].
7) **Electricity**: The rural areas have irregular or unreliable power supply issues [6], [7], [36], [48]. The rural

telecommunication system highlight the need for renewable energy sources and highly power efficient system. The 5G systems are 10x times power efficient compared to 4G systems which is an advantage for rural areas. The power supply in rural parts of India is inconsistent which needs to addressed [49].

8) **Digital awareness**: Another challenge that needs to be addressed is digital awareness among the rural users. The people need to be educated on the benefits and usage of digital solutions. When people understand the advantages of rural digital solutions and how it will improve their life, then people would demand for better services [4], [36], [50]. Another factor which needs to considered for the Indian setting is the local language in different parts of the country. The content should be available in different languages to attract subscribers on the network [51], [52]. The awareness of using digital services in improving their lives increases the value of service especially in terms of economic growth and opportunities for the younger generations in unlimited [4], [29].

## V. Possible technological solutions for Indian settings

In this section, the various possible technologies which would help in improving internet connectivity are discussed.

### A. Overview of possible technologies for rural connectivity

As previously mentioned, rural connectivity objective must overcome numerous obstacles, including innovations in technologies. Technologies such as 3G and 4G are existing in rural areas [7] but these technologies are expensive as network sharing is difficult between operators. Furthermore, passive sharing does not reduce the cost significantly as 5G and open networks [53]. Technologies such as 5G, hot air balloon base stations, drone base stations, satellite internet and backhaul, long-range WiFi networks, unmanned aerial vehicle, and TV whitespace (TVWS) are being tested for economical and easy to setup rural solutions.

In Table I, various rural technologies and their strengths as well as weaknesses are summarized. The challenge with 5G networks for rural solutions that needs to be addressed are local spectrum licensing, ease of new entrants in local markets, subscription of services, performance of network slicing. Furthermore, in [6], [41], [48], the importance of renewable energy for rural connectivity was investigated, with the issue of connection to an alternate energy source in the absence of renewable sources. Similarly, though TVWS communication has a long range, but parameter limits exist for deployment in terms of factors such as power levels, maximum allowable interference levels and spectrum licensing [6], [10], [41], [48]. The price and reliability issues associated with solutions such as hot air balloon base stations, drone base stations, satellite backhaul, long range WiFi, and unmanned aerial vehicles must be solved for these solutions to become scalable, according to the rural connectivity study [6], [41]. These solutions are not yet totally appropriate for use in rural areas at the current stage for a wide-area network. Different rural technologies are compared and their viability is assessed in terms of network architecture, performance, important characteristics, and deployment concerns in the [43].

### B. Rural trials around the world

*1) 5G and network slicing:* 5G is fifth generation of communication systems supporting three major use-cases such as enhanced mobile broadband (eMBB), ultra-low latency communications (uRLLCs), and massive machine to machine communications (mMTC). To address the issue of rural internet connectivity, researchers began testing the suitability of the design of a 5G communication infrastructure to address the needs of rural use-cases. In [38], [65], [66], the research on deploying and operating a nationwide 5G network with a duration 10 years was investigated. It was concluded that policy and technological advancement along with smart innovations in rural telecommunication business models will drive the connectivity. It was also understood that MNOs take a long time to deploy 100% coverage, because the provisioning of telecommunication services is a costly procedure with a low or negative return on investment in rural or remote rural areas. The risk of 5G adoption is also considered, along with the technological and economic viability. The use of spectrum sharing in local 5G networks deployed by the InP is explored in [61], [67], [68]. The studies highlights the benefits and ease of spectrum sharing in 5G bands especially at the higher frequency bands. In [63], the authors demonstrates the usage of 3.5 GHz band for long-range coverage to provide broadband services. The results showed the potential usage of 3.5 GHz CBRS bands in providing broadband services especially in rural scenarios.

Network slicing, which encompasses applications ranging from slicing in the radio access network to end-to-end network sharing, is one of the 5G architectures that is being investigated extensively for rural solutions. The network slicing for broadband related applications is explored in [30] to improve rural broadband services, which is backed by eMBB services of 5G systems. The research shows the discussions on bridging the digital divide issues and focusing on multi-tenancy sharing of infrastructure to make rural connectivity sustainable. In [39], the authors advocated using a single slice to give basic but free internet connectivity to people who cannot afford it in the Democratic Republic of Congo (DRC), while the rest slices are utilised for commercial operations. The solution educates rural residents about the advantages of having access to the internet. The idea of employing 5G NHN for vertical industries is being tested in a pilot study on the Orkney Islands, United Kingdom, where one of the slices will be utilised for BBC [69] broadcasting applications and other rural industrial verticals. The learning's from the trial is that it is an innovative option to encourage rural connectivity. The business models and value network configuration for 5G neutral host networks (NHN) in the rural areas is explored in [12]. The research shows that integration of business ecosystem with rural connectivity solutions to improve digital divide. Furthermore, the techno-economic feasibility of using 5G NHN in rural parts of the India is studied in [13]. The results show that it has a potential to be considered for rural connectivity in India.

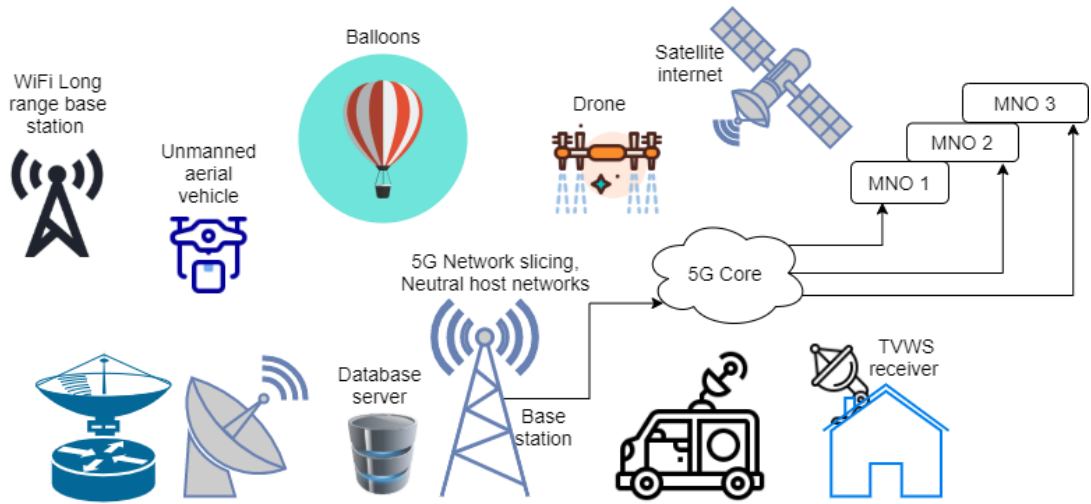

Fig. 2. Different rural technologies

TABLE I
COMPARISON OF DIFFERENT RURAL CONNECTIVITY SOLUTIONS

| Technology | References | Strengths | Weakness |
|---|---|---|---|
| 5G | [12], [13], [38], [54] | Very high speed, versatile applications, reasonable pricing, remote maintenance, software defined networking, shared spectrum, supports multiple frequencies, works well up to 500 kmph, energy efficient | High TCO, difficulty obtaining spectrum licence, requires legal approvals and network planning, revenue depends on the subscription rates |
| Satellite internet | [6], [55]–[57] | Provide coverage anywhere on the earth, high speed data rates, easy setup, supports OTT, smooth handover for mobile devices, low latency | High subscription fees, high customer premise equipment (CPE), not economically feasible, dependent on weather, no roaming for in-house antenna, lower data capacity limits compared to terrestrial networks |
| Hot air balloon/ Unmanned aerial vehicles/ Drone | [6], [58], [59] | Wireless infrastructure, long distance coverage, relatively easy to setup anywhere, short requirements for base station | Techno-economic feasibility is not fully studied, weather related challenges, legal approvals, interference planning, fuel issues, roaming agreements, continuous service requirement is not feasible |
| TVWS | [10], [37], [48], [60]–[62] | Very good coverage, supports LTE, 5G small cells and internet, Long distance transmission, wireless backhaul, long transmission ranges, lower cost | Spectrum licence, DSA technology, licence duration, middle-mile technology, user-devices do not operate on these bands, power restrictions, interference management, dynamic change in frequency of operation |
| WiFi long range | [63], [64] | Easy to setup, unlicensed bands, devices support WiFi bands, plug and play devices, in-house or local areas networks, very high data rates | Limited power supply, interference prone, customised for rural requirements, handover for mobile networks |

*2) Satellite internet:* Recently, there has been research on using low earth orbit satellites to provide internet connectivity to hard to reach rural areas and other topography areas which is not provided coverage using terrestrial telecommunication system around the world [56], [70]. Today, various satellite telecommunication companies such as OneWeb, SapceX, Eutelsat, and Blue origin are launching satellites to provide 5G and internet services for different applications [55], [57], [71]. The pricing is expected to vary depending upon the application served. For example, rural connectivity would be offered at affordable prices whereas internet connectivity for airlines and ships could be offered at higher prices. OneWeb and SpaceX aim to remove the dependency of rural areas on fibre backhaul by providing excellent coverage and data rates using satellite internet.

*3) Hot air balloon/UAV/Drone:* The application of hot air balloons/UAV/Drone is studied for digital connectivity in hard to reach areas due to topographical challenges as well as for emergency services in [6], [58], [59]. The cost of the network is highly dependent on the number of drones required to support the use-cases. The trials were performed for drone to acts as temporary base stations during network disruption in [59], [72]. The results were satisfactory in the test trials. The feasibility of drone for a longer duration of usage is yet to be studied. Companies such as Google, Facebook, Amazon and Walmart are researching on use of these technologies for 5G service provisioning [73].

*4) TVWS:* TVWS is a telecommunication technology in which information is transmitted using the TV frequency bands. After the digitization of the TV channels, a swathe of frequency became available around the world. In different test trials TVWS is studied for improving rural connectivity. In [10], [37], [48], [60], the authors study the network by using TVWS technology to provide very long distance wireless backhaul capacity. The TVWS require last-mile connectivity on a different frequency bands such as 2G, 3G, 4G, 5G or unlicensed spectrum, for operations on end-users devices. The frequency bands available for usage varies in different countries around the world. The application of TVWS in combination with WiFi is trialled in India by researchers of IEEE rural 5G broadband standards group [11]. The authors in [61], [74] highlight the different use-cases of shared spectrum and other test trials of TVWS in the UK. The results explain the benefits of using shared spectrum bands for providing connectivity and the different frequency sharing technologies around the world.

*5) WiFi long range:* This technology is being trailed as well as tested by different researchers and is easier to setup [6], [61], [63], [75]. The key advantages of long range WiFi is low cost and quick deployment but the challenges lies in the maximum permissible operational power levels as it is operates in unlicensed spectrum bands [76]. This technology also uses MIMO systems to provide long range coverage in rural areas and adhere to the usage policy of unlicensed spectrum bands. The recent study in IEEE 802.11ah shows that it has a potential for long range backhaul and long outdoor network [77] with power restrictions. The paper discussed the strengths and weakness of the technology and provide a survey of research from 2002 - 2014. This concept is extended further and researched to the networks in 2021 for topics such as fair allocation of resources, co-existence with other networks and use-case suitability.

## C. Overview of rural trials in the Indian scenario

In this section, a short discussion of the tests and trials of the technologies discussed earlier in the Indian scenario is presented. The need for sustainable internet business models and possible technologies is discussed in [78]. The rural network will only become sustainable when the business model suits the demand and capacity available for that village. The suitability and challenges of using long range WiFi, 2G/3G, LTE, and a few other technologies has been discussed in [79]. The research shows that WiFi involves lower cost and easy setup but prone to interference whereas the LTE is suitable to support roaming for rural customers though it is expensive. In [80], the authors present the 'DakNet' using long range WiFi to connect rural parts of India. The usage of LTE for rural connectivity is discussed in [7]. The author highlights the need for cheaper CAPEX cost and low spectrum licensing costs. The researchers in [75] shows the usage for long range WiFi for affordable connectivity for fishermen using the concept of network hops. The network provides coverage up to 100 km using the multi-hop network concepts. In [11], the authors discuss the technology and architecture related to using 5G IEEE standard network slicing for rural connectivity using WiFi as last mile network. This network also uses TVWS as a backhaul solution. All these researchers show the use of WiFi routers in the village for end-users to access internet. There is a need to make mobile technologies easily accessible and available for increasing the uptake of mobile services. In [15], the author highlights the need to eliminate spectrum fees as a supportive policy to encourage 5G deployments in the India scenario and in future encourage 6G networks in the rural areas especially in India.

## D. 6G - integration of different technologies

The latest generation of telecommunication systems being researched is 6G. Tetra-Hertz frequency band communication is supported by 6G and a minimum data speed of 1 Gpbs speeds per user. Furthermore, 6G is expected to offers an extremely high data rates, extraordinary features and customised network as per the design requirements. In [6], [41], [42], a survey on delivering 6G connectivity in rural areas, challenges, different technologies and viable solutions is discussed. According to the authors, 6G is expected to be a novel solution which would address rural connectivity issues around the world providing the policies are supportive for network upgradation and spectrum licensing. There are different research groups which are working on 6G research ideas to tackle digital divide issues. The 6G includes rural connectivity as one of the key use-cases unlike 5G [6]. The key rural connectivity aims of 6G as stated in [5], [6], [15], [41], [53] are as summarised below:

- Low-cost rural connectivity technology, backhaul solutions to reduce the overall cost, remote monitoring and maintenance of the network.
- Efficient spectrum usage in the spectrum sharing bands, energy consumption to address unreliable energy supply, resource allocation to support multiple applications as well as utilisation. The network supports guaranteed QoS and high reliability to encourage MNO investments for the rural connectivity ecosystem.
- Smart infrastructure for telecommunication services in rural areas.

TABLE II
5G AND 6G COMPARISON

| Feature | 5G | 6G |
|---|---|---|
| Maximum frequency | 90 GHz | 10 THz |
| Technologies | mmWave, IoT, industry4.0, network slicing, massive MIMO | Satellite integration, THz, autonomous vehicles, haptics communications, intelligent surfaces |
| End-to-end latency | 10ms | 1 ms |

The 6G rural connectivity solutions would encourage the business for micro-operators or infrastructure providers (InP), as well as the MNOs, internet service provider (ISP), temporary service provider who can lease services from the InP for providing services to their end-customers. The demand and supply should be proportional to keep the cost reasonable.

The proposed research ideas include use-cases for various stakeholders such standardisation bodies, national and international regulations, equipment manufacturers, MNOs, ISP, schools, hospitals, councils, indoor-building service providers, satellite service providers, vehicular connectivity providers and hospitals. The 6G rural solutions could also be extended to the Indian scenario.

## VI. Conclusion

This research paper provides a survey on rural connectivity challenges and possible solutions in India. The problem of providing high speed connectivity to more than 70% population of India is serious that needs to be addressed. Thus, in this paper the use-cases, the various government initiatives and projects, challenges, and technologies required for rural connectivity in the rural Indian scenario is explored, listed and analyzed. The typical solutions and its strengths as well as weakness is listed. The lessons learnt from the survey is that no single solution is suitable for all villages. The challenges, initiatives and technologies needs to be considered and weighted according to the suitability to find a solution to cater to the use-cases and demand of the rural areas. Finally, this paper presents an overview of possible rural solutions using the 6G technologies.

## Acknowledgement